\documentclass{pic2012}
\def\runauthor{Ehud Duchovni}
\def\shorttitle{Jet's substructure}

\newcommand{\pt}{$p_T$}
\newcommand{\ttbar}{$t \bar t$}
\newcommand{\ppbar}{$p \bar p$}
\newcommand{\DR}{$\Delta R$}
\newcommand{\mj}{$m_{Jet}$}
\newcommand{\kt}{$k_t$}
\newcommand{\akt}{$anti-k_t$}
\newcommand{\dij}{$d_{i,j}$}
\newcommand{\dab}{$d_{1,2}$}
\newcommand{\npv}{$N_{PV}$}

\begin{document}

\title{Substructure of Boosted Jets}

\author{Ehud Duchovni}

\address{Department of Particle Physics and Astrophysics\\
Weizmann Institute of Science\\
Rehovot, Israel\\
E-mail:ehud.duchovni@weizmann.ac.il }

\maketitle

\abstracts{Jets with transverse energy of few TeV are becoming now common in LHC data. Most of these jets are produced by QCD processes and some from the collimated decay of highly boosted objects like W, Z, $H^0$ and top-quark. The study of such QCD jets may shed light on QCD showering processes and the identification of the jets coming from decays may  test the Standard Model under extreme conditions and may also provide the first hints for Physics Beyond the Standard Model.\\
A short review of jet algorithms, Correction procedures for pile-up effects and commonly used substructure observables are described.
}

\section{Introduction}{\label{int}}
Jets are collimated showers of particles produced in high energy collisions. Vaguely speaking each well-separated out-coming parton (quark or gluon) is converted into a jet. Quantum Chromo Dynamic (QCD)  somewhat modify this simplistic picture. In run II of the Tevatron accelerator, and especially at the recent 8TeV run of LHC, one encounters jets with transverse momentum (\pt) of up to few TeV. At such a high \pt~the width (or alternatively the mass) of a QCD jet, may, at times, become quite large. The study of such highly boosted massive jets sheds light on QCD ~\cite{review} showering mechanism and provides excellent environment for perturbative QCD studies. But, in addition to QCD jet one may also expect the production of highly boosted~W, Z, $H^0$ and top quarks. The opening angle(s) between the decay products of one of these heavy objects become so small that the highly boosted object shows up as a single massive, rather than two or three, jet.  The study of such boosted particle in the framework of the  Standard Model (SM) may serve as a test of the model under extreme conditions, but, more electrifying, such a study may, according to some of the Physics Beyond the SM (BSM) scenarios, provide us with the first observations of BSM signals.
Hence, an efficient separation between highly boosted QCD jet and highly boosted heavy object can facilitate both  tests of the SM and  searches for first hints of Physics Beyond the SM. Such a separation can only be done by looking at the internal structure of these jets, namely, their sub-structure.

The challenge is made even harder due to the presence, in pp or \ppbar~collisions, of an underlying event and due to the presence of multiple interactions. The Underlying Event (UE) is the  diffuse radiation, partially coherent with the hard scatter, due to the disintegration of the remnants of the two colliding hadrons. It gives rise to a significant number of particles most of which are at small angle w.r.t. the beam direction. Multiple Interactions (MI) occur when two beams of $\approx 10^{11}$ proton cross the path of each other. In such a case few protons from one beam may collide with few protons of the other. Most of these collisions are soft yet they produce something like a cloud of particle moving out in all directions, piling up and  potentially masking the small fluctuations in energy density that may provide a clue to the jet's origin. While MI are a nuisance when looking at the sub-structure of a jet, they are inevitable when opting for high luminosity. Currently, the LHC runs with $\mu$ - the average number of MI per event  - of over 20, and the number will be increased to few hundreds at the super LHC (sLHC).

In this paper we do not intend to cover neither the topic of jet-substructure nor that of correction techniques for UE and MI. Rather, we try to provide an short description of the experimentally applies quantities and techniques. In section~\ref{alg} a very brief reminder of the classes of jet finding algorithms is presented. It is followed, in section~\ref{JCOR}, by the outlines of few algorithms that provide partial corrections to the obscuring effects of UE and MI. Then, in section~\ref{JSJSS}, a concise list of the most popular jet-shape and jet-substructure observables.

Then, in . In section 5 and jet's sub-structure, in section 6. Few concluding remarks are presented in section 7 together with a short outlook.

\section{Jet-Finding Algorithms}{\label{alg}}
Broadly speaking one can split the multitude of jet finding algorithms into two classes: \textit{Cone Algorithms} and \textit{Sequential recombination} (SR) ones.
\subsection{Cone Algorithms}
Simple intuition dictates that most of the particles belonging to a given jet are contained inside a narrow cone traced around the jet direction. In order to implement this idea one must first define a procedure to uncover the jet direction and one must predefine the opening angle of such a cone. The identification of the jet direction is  usually done by an iterative procedure, starting, e.g. from the most energetic constituent. The opening angle is a free parameter and is expressed by \DR~when \DR~between particle i and j is defined by $\sqrt{\eta_i-\eta_j)^2+(\phi_i-\phi_j)^2} $. \DR~ values between 0.2 and 1.0 are in use. In addition to \DR, one must set a lower \pt~cutoff value above which a group of particles may be consider as a jet, This $E_T^{min}$ is the second free parameter of the cone algorithms and can be adjusted by requiring that a certain percentage of the total $E_T$ is contained in jets.

\subsection{Sequential Recombination Algorithms}{\label{SRA}}
In Sequential Recombination (SR) algorithms one starts with all the constituents of the jets\footnote{when a particle can be a charged track, a calorimeter cell or any other local energy deposition} and determines a predefined \textit{similarity} between each pair. The similarity can be a measure of proximity between two such particle, e.g. the angle between them or their invariant mass. The two particles with lowest proximity are fussed together into an new particle and the process continues till all particles are merged together or till a certain criteria, e.g. maximal proximity, is reached. The \kt, Cambridge-Aachen (CA) and \akt~ algorithms are among the most popular SR procedures. The proximity is defined, for these algorithms, by $d_{i,j}=min(p_i^{2q},p_j^{2q})\frac{\Delta R_{i,j}^2}{R^2}$ where q assumes the value 1,0 and -1 respectively. R determines the 'size' of the jet and is known as the radius parameter. The \akt~algorithm traces jets with regular shape while the CA and \kt~give rise to less circular jets. One of the most popular implementation of these jet algorithms is in the framework of the fastjet package~\cite{fastjet} which is extensively used below.

\subsection{Infra Red and Collinear Stability}{\label{IR}}
A basic prerequisite for every reliable jet algorithm is that its outcome is independent of the presence of soft particles, such as produced by the UE and MI, and is independent of a possible split of a hard particle carrying a momentum $p_T^{(1)}$ into two particles carrying momentum $p_T^{(2)}$ and $p_T^{(3)}$ such that $p_T^{(1)}=p_T^{(2)}+p_T^{(3)}$. The former is known as Infra Red safety and the later as Collinear safety (IRC safety). Generally speaking the cone algorithms are IR unsafe while the SR ones are safe.  This makes the SR ones better adapted for jet-shape and substructure studies.

\section{Corrections of UE and MI effects}{\label{JCOR}}
The MI and to a certain extent also the UE, may mask some of the interesting features of jet's substructure. Quite a few techniques has been proposed to cope with this obstruction. Most of these techniques relay on the fact that the energy density due to MI and UE is relatively low and uniform. A brief description of several such techniques is given below:

\subsection{Trimming}{\label{Trimming}}
In this procedure~\cite{trimming} one applies a SR jet algorithm (usually \kt~or CA) with  a small radius parameter ($R_{sub}$) to the jet constituents. This results in a set of sub-jets and those with low \pt, namely, $p_T^i/p_T^{Jet}<f_{cutoff}$ are removed from the original jet leaving only the high energy density spots. Two free parameters, $R_{sub}$ and $f_{cutoff}$ have to be optimized for each study. Typical values for $f_{cutoff}$ and  $R_{sub}$ are 0.03 and 0.2 respectively.

\subsection{Pruning}{\label{Pruning}}
Here~\cite{pruning} one applies a SR jet algorithm while rejecting jet constituents which have low \pt, and are isolated from other jet constituents, namely:
if $\frac{p_T^i}{p_T^i+p_T^j}<Z_{cutoff}$ or $\Delta R_{i,j}>R_{cutoff}$ one refrains from merging these sub-constituents with the main jet. As for Trimming, also while Pruning one has to optimize two parameters: the $Z_{cutoff}$ and the $R_{cutoff}$. Typical value for $Z_{cutoff}$ is 0.1, while $R_{cutoff}$ is determined on event-by-event basis and is taken to be $R_{cutoff} =\alpha \frac{m_{jet}}{p_T^{jet}}$ and $\alpha$ is close to one.

\subsection{BDRS Filtering}{\label{Filtering}}
This is slightly more elaborated method~\cite{BDRS}.  One backs off one step in the SR jet algorithm that has been applied to define the jets. Each of the two sub-jets is characterized by its \pt~and its mass $m_{i_1} $ and $m_{i_2}$ where by construction  $m_{i_1} $ is the heavier one. One then defines the mass asymmetry parameter $\mu=\frac{m_{i_1}}{m_{jet}}$ (mass drop). One also defines the momentum asymmetry parameter  $y_{cutoff}=\frac{min(p_T^{(i_1)},p_T^{(i_2)})}{m^2_{jet}}\Delta R^2_{i_1,i_2}$. A jet is deemed to be \textit{clean} if the mass drop is large and the momentum asymmetry is moderate. Should one of these conditions violated, the less massive jet, $i_2$, is assumed to be an artifact of the pileup, is removed and the procedure is repeated on jet $i_1$. Typically $\mu \approx0.7$ and $y_{cutoff} \approx 0.1$ are used.\\

\begin{figure}[htbp] %
   \centering
   \includegraphics[width=11cm]{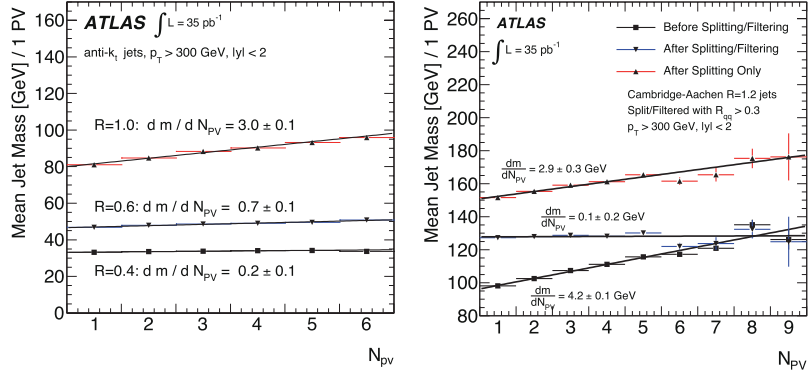}
   \caption{The mean mass for jets with $p_T > 300$ GeV as a function of \npv. Comparisons show the effect for \akt~jets with different R-parameters (left) and
 Cambridge-Aachen R = 1:2 jets with and without splitting and filtering procedure (right). Each set of points is fitted with a straight line.}
   \label{fig:filteredmass}
\end{figure}

The distortion of the various substructure quantities should increase with the increase of the number of MI which is well approximated by the number of primary charged-tracks vertices \npv. The average jet's mass, for example, should be higher for events having \npv$>>1$ than that measured for events having \npv=1. An efficient correction procedure should correct for this effect and render the average jet's mass independent of \npv. The measured average jet mass before and after the  correction procedure described in this subsection is shown in Figure~\ref{fig:filteredmass}~\cite{atlas.sub}.

\subsection{Complementary Cone Method}
\begin{figure}[!htbp]
\begin{center}
\includegraphics*[width=11.0cm]{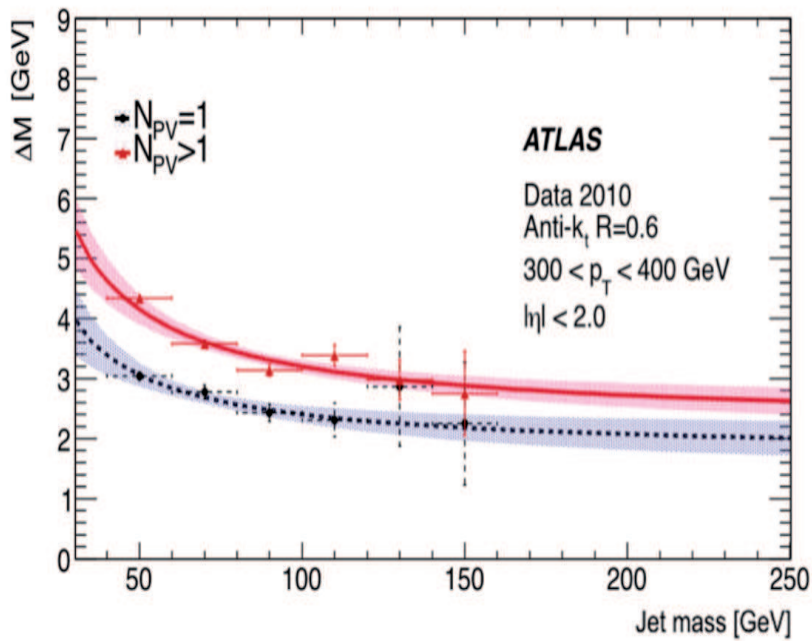}
\end{center}
\caption{The size of the correction to the jet's mass for the \akt~algorithm with R=0.6.}
\label{fig:masscor}
\end{figure}

{\label{Complementary}}
A complementary cone is drawn~\cite{compcone} at a right-angle in azimuth to the jet $\phi_{comp}=\phi_{jet} \pm \frac{\pi}{2}, ~\eta_{comp}=\eta_{jet}$ and the energy deposits in this cone are added into the jet such that the effect on each of the jet properties can be quantified. The shift in each observable after this addition is attributed to pileup and the underlying
event (UE). The effects of these two sources are separated by comparing events with \npv = 1 (UE only) to those with $N_{PV} > 1$ (UE and pileup). The presence of additional energy in events with $N_{PV} > 1$ affects the substructure observables in different ways~\cite{Alon}. For example, the shift in jet's mass is given by: $\Delta m=p_{0_M}+\frac{p_{1_M}}{M}$ where $p_{i_M}$ are coefficients hat has to be determined from the data themselves. One should, however, bare in mind that this approximation is valid only when $\Delta M<<M$. The accuracy of this prediction is shown in  Figure~\ref{fig:masscor} the \akt~algorithm with R=0.6 is applied and jets with $300<p_T<400$GeV are selected.

\section{Jet Shape and Jet Substructure observables}{\label{JSJSS}}
The energy density inside a jet is usually higher near the jet axis dropping toward the boundaries of the jet. Quantities that try to quantify this property are usually referred to as \textit{Jet Shape} or \textit{substructure} observables. The simplest, which is overlooked as a substructure quantity, is simply the mass.

\subsection{The Jet Mass}
\begin{figure}[!htbp]
\begin{center}
\includegraphics*[width=11.0cm]{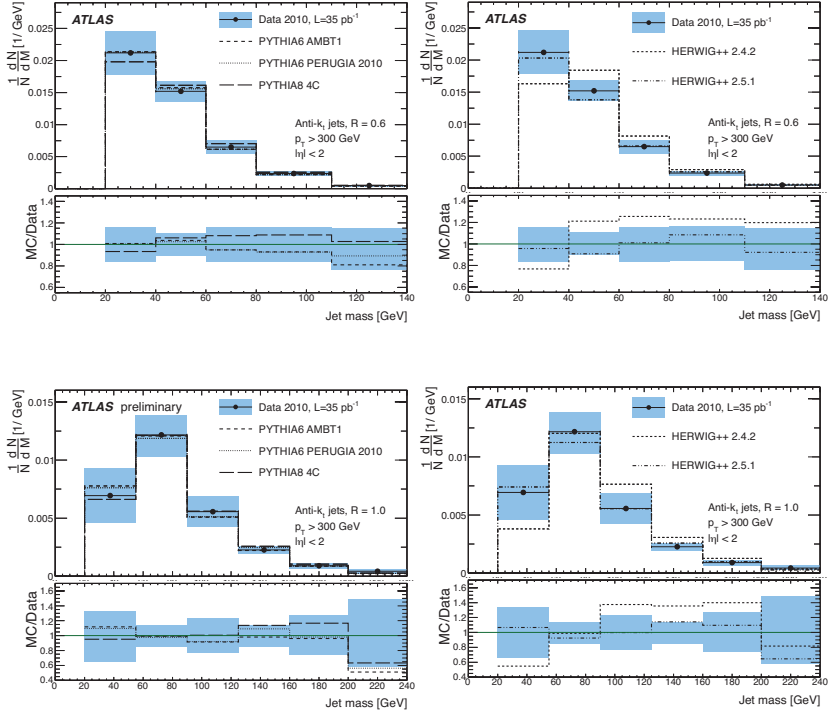}
\end{center}
\caption{The mass distribution for the \akt~algorithm with R=0.6 (top) or 1.0 (bottom)  jets having \pt$>300$ GeV in $|\eta|<2$ before (red) and after (black) the mass correction. Also shown (blue) is the mass distribution for $N_{PV}=1$ events . The distributions are compared to Pythia (left) and Herwig (right) predictions.}
\label{fig:massdist}
\end{figure}

{\label{Jmass}}
The mass distribution of boosted jets is a simple example for a jet substructure observable. QCD predicts that the more massive  jets acquire most of their mass through a single gluon bremsstrahlung in which the energy of the original parton is equally partitioned between the leading particle and the main emitted gluon. Under some simplifying assumptions an absolute prediction of both size and shape of \mj~distribution is computed~\cite{jet-mass} and given by:
\begin{equation}
\frac{d\sigma(R)}{dp_T dm^{jet}}=\sum_{q,g}J^{q,g}(m^{jet},p_T,R) \frac{d\hat \sigma ^{q,g}(R)}{dp_T}
\end{equation}
where
\begin{equation}
J^{q,g}(m^{jet},p_T,R) \approx \alpha_s(p_T) \frac{4C_{q,g}}{\pi m^{jet}}log(\frac{Rp_T}{m^{jet}})
\end{equation}
and $C_{q,g}$=4/3 for quarks and 3 for gluons.
The measured jet mass distribution is presented in Figure~\ref{fig:massdist}

\subsection{Width}
\begin{figure}[!htbp]
\begin{center}
\includegraphics*[width=11.5cm]{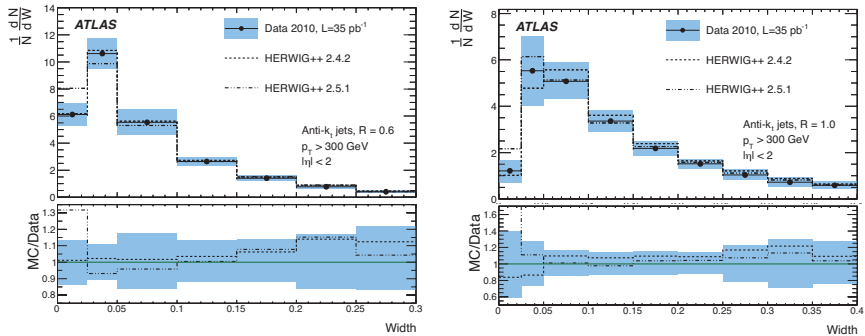}
\end{center}
\caption{The jet's width distribution for the \akt~algorithm with R=0.6 before (red) and after (black) the mass correction. Also shown (blue) is the mass distribution for $N_{PV}=1$ events. The measurement is compared with Herwig predictions}
\label{fig:widthdist}
\end{figure}

{\label{width}}
The width of a jet is defined by
\begin{equation}
W_{jet}=\frac{\sum_i \Delta R^i p_T^i}{\sum_i p_T^i}
\end{equation}
where $\Delta R^i=\sqrt{(\Delta \eta^i)^2+(\Delta \phi^i)^2}$ is the radial distance between the $i^{th}$ component of the jet and the jet direction.
The width distribution is shown in Figure~\ref{fig:widthdist}

\subsection{Eccentricity}
\begin{figure}[!htbp]
\begin{center}
\includegraphics*[width=11.0cm]{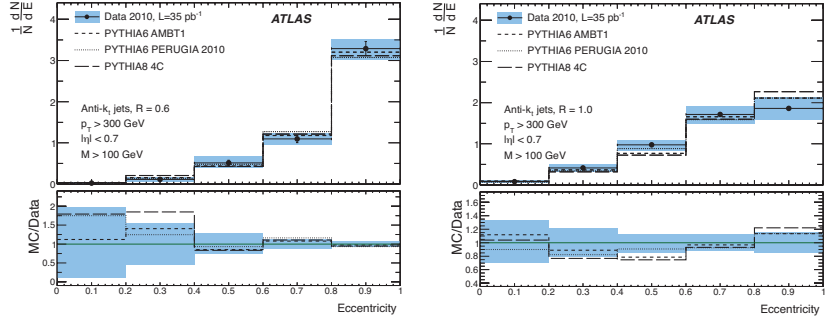}
\end{center}
\caption{The jet eccentricity distributions for high mass (M > 100 GeV) leading jets defined with the \akt~algorithm with R = 0.6 (top) and R = 1.0
(bottom) jets in the full 2010 ATLAS data set, corrected for pileup to reflect the particle level.}
\label{fig:eccentricity}
\end{figure}

{\label{eccentricity}}
The jet eccentricity, E, is calculated using a principal component analysis (PCA)~\cite{PCA}. The PCA method provides the vector which best describes the energy-weighted
geometrical distribution of the jet constituents in the ($\eta,\phi$) plane. The eccentricity is used to characterize the deviation of the jet profile from a perfect circle in this
plane, and is defined as:
$$\epsilon=1=\frac{v_{min}}{v_{max}}$$
where $v_{min}$ and $v_{max}$ are  the maximal and minimal value of the variance of the jet constituents' positions with respect to the principal vector.
The distribution of eccentricity is depicted in Figure~\ref{fig:eccentricity}. This is one of the very few sub-structure quantities that are sensitive to non-isotropic energy flow in the plane perpendicular to the jet axis.

\subsection{Planar Flow}{\label{width}}
 The planar flow~\cite{planar} measures the degree to which the jet's energy is evenly spread over the plane across the face of the jet (high planar flow) versus spread
linearly along the event axis (small planar flow). The planar flow is defined by:
\begin{equation}
P=4 \times \frac{det(I_E)}{Tr(I_E)^2}=\frac{4\lambda_1 \lambda_2}{(\lambda_1 + \lambda_2)^2}
\end{equation}
and the (i,j) components of the two dimension momentum tensor $I_E$ are defined by:
$$I_E^{i,j}=\frac{1}{M} \sum_k \frac{1}{E_k}p_{k,i}p_{k,j}$$
where the summation is over all of the jet constituents and $E_k$ is the energy of the constituent and
$p_{k,i}$ \&  $p_{k,j}$ are the $i^{th}$  and $j^{th}$ momentum components in the $(x,y)$ plane perpendicular w.r.t. the jet axis.

Very small planar flow values corresponds to two back to back jets event topology. isotropic energy density corresponds to planar flow value of unity.

Jets with many body kinematics are expected to have a planar flow distribution that peaks slightly below unity. This is the situation for low mass jets where the mass is built by emission of many soft gluons isotropically around the jet axis.Highly boosted massive jets  are well-described by a unique hard gluon emission and, therefore, should give rise to jets with very low value of planar flow~\cite{boostedA}.

\begin{figure}[!htbp]
\begin{center}
\includegraphics*[width=11.0cm]{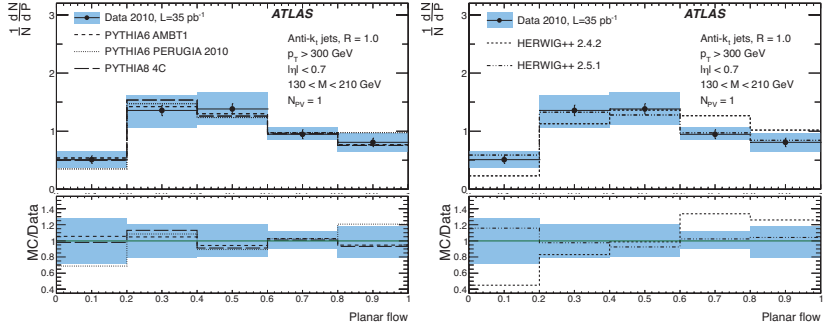}
\end{center}
\caption{The jet planar flow distributions for high \pt~($p_T>300$ GeV) massive jets ($130 < M < 210$ GeV),The \akt~with R=1.0 have been used and only events with $N_{PV} = 1$ are included. The distribution was corrected for detector effects}

\label{fig:planar}
\end{figure}

The planar flow distribution for jets with mass consistent with that of the to-quark ($130<M_{jet}<210$ GeV) is shown in Figure~\ref{fig:planar} where jets have been reconstructed with the \akt~algorithm with R=1.0 and after the application of a $p_T>300$ GeV cut.

\subsection{Angularity}{\label{angularity}}
The angularity, defined by:
\begin{equation}
\tau_a=\frac{1}{M} \sum_i E_isin^a\theta_i[1-cos \theta_i]^1-a
\end{equation}
for small angles it is approximated by:
\begin{equation}
\tau_a=\frac{2^{a-1}}{M} \sum_i E_i \theta_i^{2-a}
\end{equation}
is yet another quantity that measures the shape of the energy density distribution around the jet axis. It is IR safe for $a\le2$. usually only the $a=-2$ moment ($\tau_{-2}$)  is studied.
$\tau_{-2}$ is can discriminates between QCD jets and jets originating from boosted heavy particle decays sinceQCD is expected to have a broader $\tau_{-2}$ tail~\cite{angul}.

The angularity of jets with two-body kinematics should peak around a minimum value$\tau_a^{peak} \approx (\frac{M}{2p_T})^{1-a}$, which corresponds to the situation in which the
two hard constituents are in a symmetric \pt~configuration around the jet axis. An estimate for the maximum value of  $\tau_{-2}$ is obtained by considering the situation by which the jet contains a hard, near axis, component and a soft, near the edge, component. In this situation (small angle radiation)  $\tau_{-2}^{max}=(\frac{2}{R})^a\frac{M}{2p_T}$.

QCD makes explicit prediction for the shape of $\tau_{-2}$ and can, therefore, be tested by focusing on jets with $100<M^{jet}<130$ GeV which is situated between the W/Z and top-quark regions. The $\tau_{-2}$ distribution is shown in Figure{fig:angul}

\begin{figure}[!htbp]
\begin{center}
\includegraphics*[width=11.0cm]{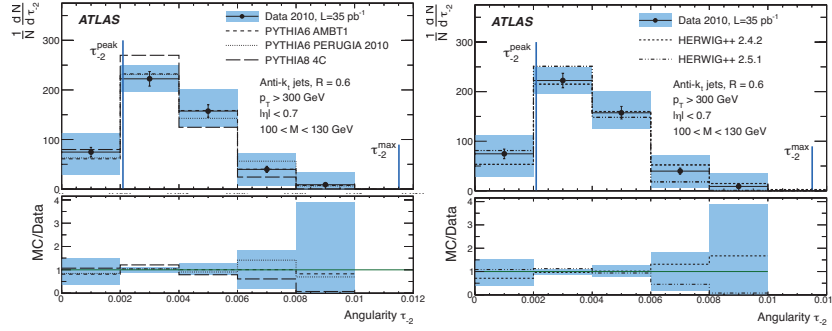}
\end{center}
\caption{The angularity distributions for high mass ($100 < M < 130 $ GeV), high \pt~ ($p_T>300$  GeV) \akt~jets with  R = 0.6, corrected to particle level. The peak and maximum positions predicted by the small angle approximation are indicated.}
\label{fig:angul}
\end{figure}

\subsection{\kt~ Splitting Scale}
\begin{figure}[!htbp]
\begin{center}
\includegraphics*[width=11.0cm]{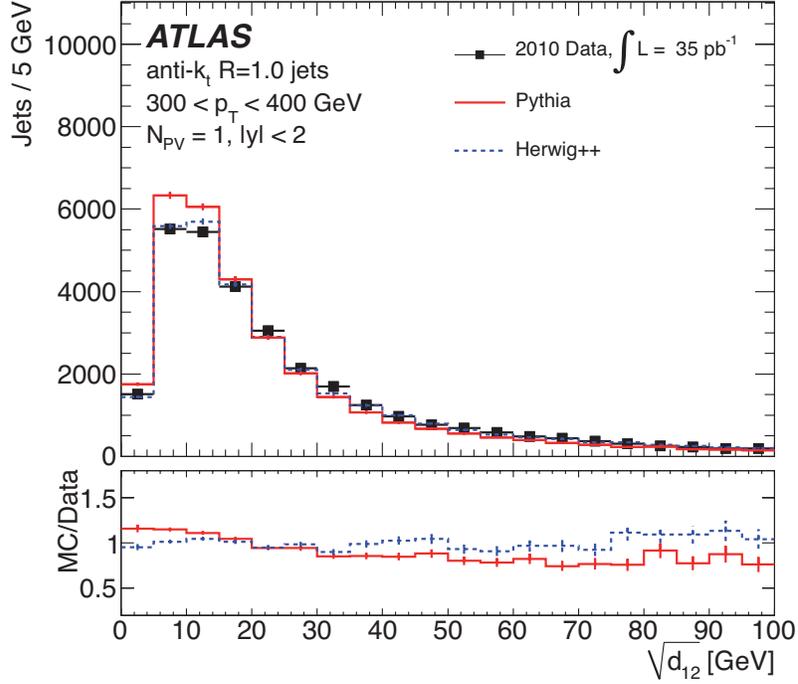}
\end{center}
\caption{The \kt~splitting parameter \dab~ for jets with $|\eta|<2.0$ and with $300<p_T<400$ GeV.}
\label{fig:d12}
\end{figure}

{\label{splitting}}
A coarse look into the jet's substructure is provided by the fist \kt~splitting scale. In the framework of this jet algorithm the order by which constituents are merged is determined by the \textit{distance}, \dij~between them. The further they are the later they get merged. This distance is defined by:
\begin{equation}
d_{i,j}^2=min(p_T^{(j_1)},p_T^{(j_2)}) \Delta R_{i,j}
\end{equation}
At the last stage of jet formation, the two last constituents (usually the outcome of the merger of many) are as a distance that is denoted by $d_{1,2}$ from each other~\cite{dij}.  Had the jet been the outcome of a two-body decay of a heavy boosted particle (e.g. $H^0 \rightarrow b \bar b$, the last merger would usually be between the jets originating from the decay products of that particle, namely the b-jets.  QCD jets  tend to give rise to  $d_{1,2} \approx 1/10$ alas with a long tail to higher values while a heavy particle will yield  $d_{1,2}$ in the vicinity of m/2. Hence, $d_{1,2}$ may be used to identify such a situation. The distribution of \dab~is shown in Figure~\ref{fig:d12}

\subsection{N-subjettiness}
\begin{figure}[!htbp]
\begin{center}
\includegraphics*[width=11.0cm]{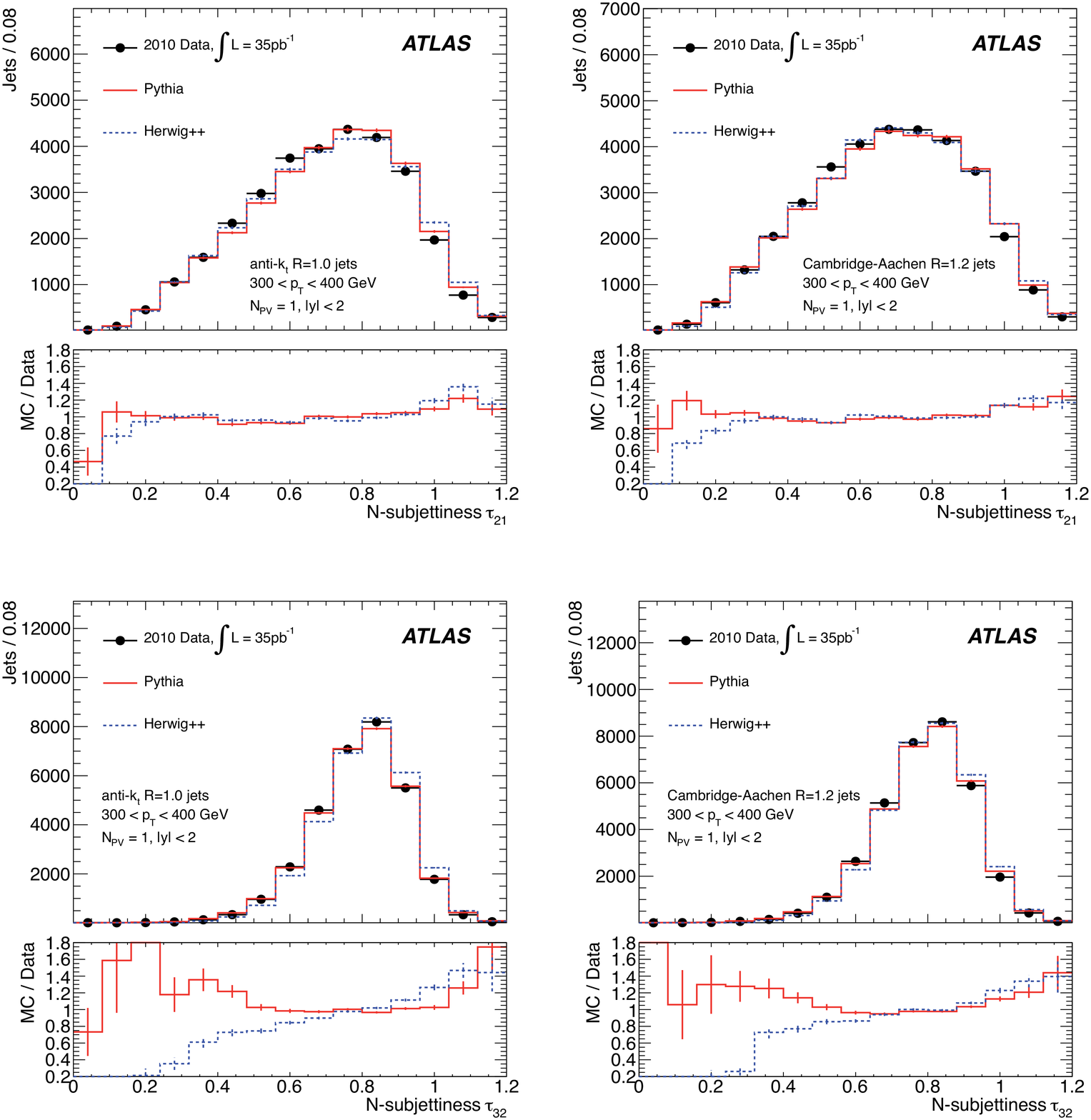}
\end{center}
\caption{The distributions of $\tau_{12}$ and $\tau_{32}$.}
\label{fig:taurat}
\end{figure}

{\label{jettiness}}
A finer look into the substructure of jets is provided by the N-subjettiness~\cite{jettines}, $\tau_N$. Jets originating from boosted top quarks contain, in case the top decays hadronically, three jets. The third subjettiness  moment, $\tau_3$ sort of \textit{measures} the agreement between the jet substructure and this hypothesis. Similarly, low  $\tau_N$ values imply that the jet seems to be consistent with the assumption of being made out of N sub-jets. The subjettiness is defined by:
\begin{equation}
\tau_N=\frac{1}{d_0} \sum_k p_T^{(k)} \times min(\Delta R_{1,k}, \Delta R_{2,k}.....,\Delta R_{N,k}) \\
\end{equation}
where $\Delta R_{i,k}$ is the distance between constituent k and sub-jet i.  $d_0= \sum_K p_T^{(k)} R$, and R is for normalization purposes and is the original R-parameter of the jet algorithm.  The ratios $\tau_{12}=\tau_2/\tau_1$ and  $\tau_{32}=\tau_3/\tau_1$ are excellent discriminators between highly boosted hadronic W bosons  and top-quarks~\cite{tratio}. The distributions of $\tau_{12}$ and $\tau_{32}$ are presented in figure~\ref{fig:taurat}.

\subsection{CMS Top-Tagging Algorithm}
\begin{figure}[!htbp]
\begin{center}
\includegraphics*[width=11.0cm]{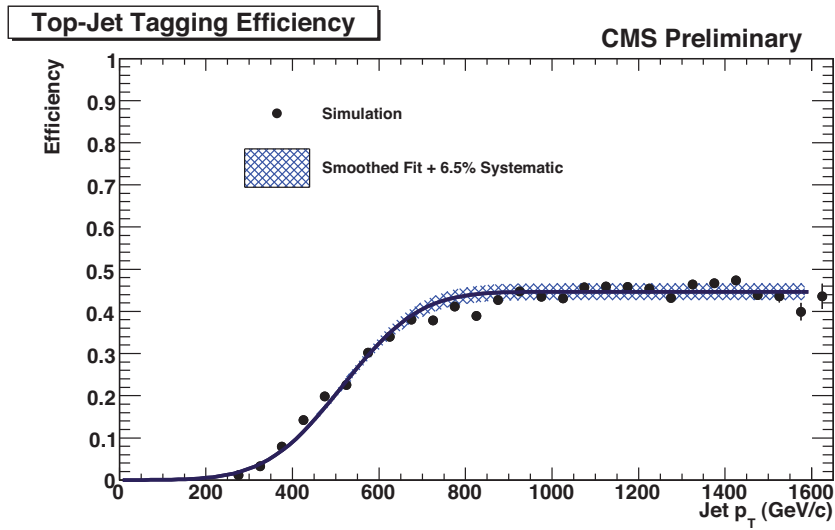}
\end{center}
\caption{The fake rate (left) and efficiency (right) obtained by the top-tagger algorithm.}
\label{fig:ttag}
\end{figure}

{\label{ttagger}}
As mentioned before, one of the practical aspects of the study of jet's substructure is to distinguish between boosted top-originated jets and the high mass tail of QCD jets. This issue has been addressed, for example, by the top-tagging algorithm constructed by CMS~\cite{ttagger}.
\begin{itemize}
\item
The algorithm accepts jets that have been reconstructed by the CA algorithm with radius parameter of 0.8 provided their \pt~exceeds 250 GeV, their mass lies between 100 and 250GeV and that their $|\eta|$ is below 2.5.
\item
{\bf primary decomposition} The parent jet's clusters are then inspected in order to determine if the jet can meaningfully be decomposed into two clusters. This is done by probing the history of the CA jet reconstruction, and by demanding that both clusters satisfy  $p_T^{cluster}>0.05 \times p_T^{jet}$. If  one of the clusters fails this minimal \pt~ requirement it is discarded and the decomposition is repeated on the other cluster till a successful decomposition is achieved, or till both constituents fail this condition.
\item
{\bf Secondary decomposition} The decomposition process is continued of each of the two parent clusters (denoted by A and B), yet the required \pt~is still compared with the \pt~of the CA jet. Only jets that have successfully decomposed into {$A,A',B,B'$}, {$A,A',B$} and {$A,B,B'$} are maintained and the group of sub-clusters is refereed to as \textit{subjets}.
\end{itemize}
The three highest \pt~subjest are assumed to come from a fully hadronic top=quark decay. They are pairs in all six possible ways and the minimal mass of these pairs is required to exceed 50GeV. Since the minimal mass usually equals, for signal events, that of the W, this last cut is very important in improving the purity of the sample.  The fake rate and purity achievable by such an algorithm are shown on Figure\ref{fig:ttag}.

\section{Conclusion}{\label{int}}
LHC opened a wide door into the exciting topic of  jet's substructure. Many new concept, like jet area, catchment - to name a few, have recently been introduces,  and will affect the experimental way of jet analysis. Here we tried to give a concise summary, and certainly not a complete one, of the concepts and techniques that are now in use. With the $\approx25fb^{-1}$ now on disk and with the 13-14TeV run in sight this field is bound to flourish.

\section*{Acknowledgements}The author would like to thank the Israel Science foundation, Minerva and Benoziyo center for High energy for their support.

\section*{Appendix}
This is place for Appendix, if any.

\end{document}